\documentclass[12pt,preprint]{aastex}
\def\gtwid{\mathrel{\raise.3ex\hbox{$>$\kern-.75em\lower1ex\hbox{$\sim$}}}}
\def\ltwid{\mathrel{\raise.3ex\hbox{$<$\kern-.75em\lower1ex\hbox{$\sim$}}}}
\def\\{\hfil\break}
\def\ie{{\it i.e.\ }}
\def\eg{{\it e.g.\ }}
\def\etal{{\it et al.\ }}

\begin{document}
\title{Optimal Moments for the Analysis of Peculiar Velocity Surveys}
\vskip 0.5cm
\author{\bf Richard Watkins\footnote{rwatkins@willamette.edu}}
\affil{\it Department of Physics, Willamette University\\ Salem, OR
97301}
\vskip 0.3cm
\author{\bf Hume A. Feldman\footnote{feldman@ku.edu},
        Scott W. Chambers\footnote{willc@kusmos.phsx.ukans.edu}, Patrick
        Gorman\footnote{gorman@kusmos.phsx.ukans.edu}, Adrian
        L. Melott\footnote{melott@kusmos.phsx.ukans.edu}}
\vskip 0.3cm
\affil{{\it Department of Physics \& Astronomy, University of Kansas\\
Lawrence, KS 66045}}

\baselineskip 12pt plus 2pt

\begin{abstract}
We present a new method for the analysis of peculiar velocity surveys
which removes contributions to velocities from small scale, nonlinear
velocity modes while retaining information about large scale motions.
Our method utilizes Karhunen--Lo\`eve methods of data compression to
construct a set of moments out of the velocities which are minimally
sensitive to small scale power.  The set of moments are then used in a
likelihood analysis.  We develop criteria for the selection of moments,
as well as a statistic to quantify the overall sensitivity of a set of
moments to small scale power.  Although we discuss our method in the
context of peculiar velocity surveys, it may also prove useful in other
situations where data filtering is required.

\end{abstract}

\noindent{\it Subject headings}: cosmology: distance scales -- cosmology: large
scale structure of the universe -- cosmology: observation -- cosmology:
theory -- galaxies: kinematics and dynamics -- galaxies: statistics

\baselineskip 23pt plus 2pt

\section{INTRODUCTION}
\label{sec-intro}

Peculiar velocity surveys are an important tool for probing the mass
distribution of the universe on large scales. In the analysis of these
surveys, galaxies or clusters of galaxies are assumed to be tracers of
the matter velocity field, which in linear theory is directly related to
the density field.  Thus peculiar velocity data can complement other
measures of the mass distribution by placing constraints on the
properties of the density field, for example, the power spectrum of
fluctuations.  Peculiar velocities also provide a powerful test of the
gravitational instability theory of structure formation.

In practice, the use of peculiar velocities to constrain properties of
the density field is complicated by several factors.  First and foremost
is the fact that a direct relationship between velocity and density
fields holds only in linear theory; this necessitates that we focus on
large enough scales so that linearity can be reasonably assumed.  This
also requires that we can adequately separate large--scale contributions
to the velocity field from small--scale, nonlinear contributions.

One of the most straightforward methods of analyzing peculiar velocity
data is to examine the statistics of low--order moments of the velocity
field, for example, the bulk flow
\citep{LP,RPK}.  The idea here is that in calculating low--order
moments the small scale modes will be averaged out, so that the values
of these moments will reflect only large--scale motion.  It has been
shown, however, that the sparseness of peculiar velocity data can lead
to small--scale modes making a significant contribution to low--order
moments through incomplete cancellation \citep{fw94,fw98}, an effect
which up to now has not been quantified.  Another drawback of this
approach is that it utilizes only a fraction of the available
information.

An alternative method is to perform a likelihood analysis using all of
the velocity information \citep{jk95}.  An obvious danger here is that
retaining small--scale, nonlinear contributions to the velocities can
lead to unpredictable biases which can skew the results
\citep{C&E}.  This method also has the disadvantage of becoming
unwieldy for surveys larger than about a thousand objects.  While
advances in computing will make this less of a problem in the future,
clearly a less time--intensive method is desirable.

In this paper we describe a new method for the analysis of peculiar
velocities which is designed to separate large and small scale velocity
information in an optimal way.  The method utilizes Karhunen--Lo\`eve
methods of data compression to construct a set of moments out of the
velocities which are minimally sensitive to small scale power; these
moments can then be used in a likelihood analysis.  Overall sensitivity
of the set of moments to small scales is quantified, and can be
controlled through the number of moments retained in the analysis.
Since the number of moments kept is typically much smaller than the
number of velocities in the survey, this method has the added advantage
of being much more efficient than a full analysis of the data.

Karhunen--Lo\`eve methods \citep{KK,KS} have recently become popular in
cosmology. A general discussion of their use in the analysis of large
data sets was done by \citet{TTH}.  In addition, Karhunen--Lo\`eve
methods have been applied to the Las Campanas Redshift Survey
\citep{matsubara00}, to velocity field surveys
\citep{hz00,silberman01}, and to the decorrelation of the power
spectrum \citep{H00,HT00}. Although we use the same general method, our
take on the formalism is quite different.  Taking advantage of the
compression techniques and the Fisher information matrix \citep{fisher},
we filter out small--scale, nonlinear velocity modes and retain only
information regarding the large--scale modes.

This paper is organized as follows. In Sec.~\ref{sec-like} we review
likelihood methods for the analysis of peculiar velocities.  In
Sec.~\ref{sec-compress} we discuss methods of data compression.  In
Sec.~\ref{sec-select}, we describe criteria for the selection of a set
of optimal moments.  In Sec.~\ref{sec-pow} we describe the power
spectrum model that will be used for our analysis, and in
Sec.~\ref{sec-anal} we discuss the application of our method to peculiar
velocity data.  In Sec.~\ref{results} we show results from performing
our analysis on simulated catalogs that illustrate the effects of
small--scale, nonlinear power and the effectiveness of our method of
analysis in removing these effects.  Finally, in Sec.~\ref{sec-conc} we
summarize and discuss our results.

\section{LIKELIHOOD METHODS FOR PECULIAR VELOCITIES}
\label{sec-like}

Several studies have used likelihood methods for the analysis of
peculiar velocity data (see, e.g., \citet{Kaiser88}).  Here we review
the most straightforward analysis of this type; one that works directly
with the observed line--of--sight peculiar velocities.  Suppose that we
are given a set of $N$ objects with positions ${\bf r}_{i}$ and
line--of--sight peculiar velocities $v_i$.  We assume that the observed
velocity $v_{i}$ is of the form
\begin{equation}
        v_{i} = {\bf v}({\bf r}_{i})\cdot {\bf \hat r}_{i} + \delta_{i}
\label{vo}
\end{equation}
where ${\bf v}({\bf r}_{i})$ is the fully three--dimensional linear
velocity field and $\delta_{i}$ is a Gaussian random variable accounting
for the deviation of a galaxy's measured velocity from the predictions
of linear theory.  We shall model $\delta_{i}$ as having variance
$\sigma_{i}^{2}+\sigma_{*}^{2}$, where $\sigma_{i}$ is the particular
observational error associated with the $i$th object and $\sigma_{*}$
accounts for contributions to the velocities of all of the galaxies in
the survey arising from nonlinear effects as well as from the components
of the velocity field that have been neglected in the linear model
\citep{Kaiser88}.  With these assumptions, the covariance matrix
$R_{ij}= \langle v_{i}\ v_{j}\rangle $ takes the form
\begin{equation}
        R_{ij} = R^{(v)}_{ij} + \delta_{ij}\, (\sigma_{i}^{2}+
        \sigma_{*}^{2})
\label{Rij}
\end{equation}
where $R^{(v)}_{ij} = \langle {\bf v}({\bf r}_{i})\cdot {\bf \hat r}_{i}
\ \ {\bf v}({\bf r}_{j})\cdot {\bf \hat r}_{j}\rangle $.
In linear theory, the covariance matrix $R^{(v)}$ can be written as an
integral over the density power spectrum
\begin{equation}
R^{(v)}_{ij} = {1\over (2\pi)^3}\int P_{(v)}(k)W^2_{ij}(k)\ d^3k =
{H^2\Omega_o^{1.2}\over 2\pi^2}\int P(k)W^2_{ij}(k)\ dk,
\label{Rvij}
\end{equation}
where $W^2_{ij}(k)$ is a tensor window function calculated from the
positions and velocity errors of the objects in the survey (for more
details see \citet{fw94,wf95}).  The derivation of this formula as well
as the definition of $W^2_{ij}(k)$ is given in appendix A.

Given the covariance matrix, we can construct the probability
distribution for the line--of--sight peculiar velocities
\begin{equation}
L(v_1,...,v_N; P(k))= \sqrt{|R^{-1}|}\ \exp\left(\sum_{i,j=1}^N -v_i
R^{-1}_{ij} v_j/2\right)
\label{like}
\end{equation}
Alternatively, if we are given a set of velocities $(v_1,...,v_N)$, we
can view $L(v_1,...,v_N;P(k))$ as the likelihood functional for the
power spectrum $P(k)$.  Typically, the power spectrum is parameterized
by a parameter vector ${\bf\Theta} = (\theta_1,...,\theta_m)$; then
$L(v_1,...,v_N;{\bf\Theta})$ becomes a likelihood function for the
parameters $ {\bf\Theta}$.  The value of the parameter vector that
maximizes the likelihood is known as the maximum likelihood estimator,
${\bf\Theta}_{ML}$.

Suppose that the true value of the parameters are given by ${\bf\Theta}=
{\bf \Theta}_o$.  The maximum likelihood estimate ${\bf\Theta}_{ML}$
will vary over different realizations of the line--of--sight velocities
$v_1,...,v_N$; we can characterize this variation with the means
$\langle ({\bf\theta}_{ML})_i \rangle$ and the variances
$\Delta(\theta_{ML})_i^2 = \langle (\theta_{ML})_i^2\rangle -\langle
(\theta_{ML})_i\rangle^2$. It has been shown \citep{KS} that in the
limit of a large number of objects, i.e. $N\rightarrow\infty$, the
maximum likelihood estimator is the best possible estimator of
${\bf\Theta}_o$ in that it is unbiased, $\langle
{\bf\theta}_{ML}\rangle= {\bf \Theta_o}$, and has the minimum possible
variances. These minimum variances are given
$\Delta(\theta_{ML})_i=1/\sqrt{F_{ii}}$, where $F_{ii}$ are the diagonal
elements of the Fisher information matrix, defined by
\begin{equation}
F_{ij}= -\langle {\partial^2 (
\ln L)\over\partial\theta_i\partial\theta_j}\rangle
\label{fishermat}
\end{equation}
evaluated at ${\bf\Theta}={\bf\Theta}_o$.  Having an estimator that is
unbiased and whose variances are characterized in terms of the Fisher
matrix simplifies our analysis considerably.  For the remainder of this
paper we shall assume that the large $N$ limit applies.

The result that the maximum likelihood estimator ${\bf\Theta}_{ML}$ is
unbiased, however, also assumes that the velocity field is Gaussian and
that the power spectrum can be well described by the given
parameterization, usually one derived from linear theory.  The collapse
of small--scale, nonlinear density perturbations can cause both of these
assumptions to be violated, and can result in ${\bf\Theta}_{ML}$ being
biased in an unpredictable way \citep{C&E}.  In order to recover an
unbiased estimator, we shall utilize methods of data compression to
filter out information about small--scale nonlinear velocities and
retaining information about large scales where the linear and Gaussian
approximations should remain valid.  While these methods are typically
used to reduce the size of an unwieldy data set without the loss of
information, here we are instead interested in using data compression as
a filter of unwanted information.

Given the difficulty in treating the general case, in the following we
retain the model of a Gaussian velocity field.  The assumption here is
that the primary effect of the collapse of nonlinear perturbations is
the modification of the power spectrum on small scales and that
departures from Gaussianity are small enough not to effect our analysis.
We will return to this issue in Sec. ~\ref{results} and Sec. ~\ref{sec-pow}.

\section{DATA COMPRESSION}
\label{sec-compress}

For a given set of velocity data, the simplest form of data compression
involves replacing $N$ original line--of--sight velocities
$(v_{1},\ldots,v_{N})$, with $N^{\prime}$ moments,
$(u_{1},\ldots,u_{N^{\prime}})$, where $N^{\prime}\le N$ (for a more
detailed discussion of data compression see \citet{TTH}).  In this paper
we will concentrate on {\it linear} data compression, where the moments
can, in general, be written as linear combinations of the velocities;
\[
u_{i}= \sum_{j=1}^N B_{ij}\ v_{j}\ ,
\] 
where $B_{ij}$ is a constant $N^{\prime}\times N$ matrix.  If the number
of moments $N^{\prime}$ is less than $N$, then replacing the $v_{i}$
with the $u_{i}$ will necessarily lead to a loss of information.
However, by a proper choice of the matrix $B_{ij}$, we can arrange it so
that the information lost is primarily associated with scales where
nonlinear effects are likely to have caused deviations from linear
theory.  Thus the process of data compression can be used to produce a
set of moments which are much less sensitive to nonlinear effects than
the original line--of--sight velocities but that still retain the
desired information about large scale power.

For simplicity, consider a model for the power spectrum in which the
power on nonlinear scales is proportional to a single parameter
$\theta_{q}$ (we will discuss a specific model of a power spectrum of
this type below).  Given a set of line--of--sight velocities,
$v_{1}\ldots v_{N}$, we can determine the value of $\theta_{q}$ within a
minimum variance of $\Delta\theta_{q}^{2} = 1/F_{qq}$, where $F_{qq}$ is
the $qq$th element of the Fisher Matrix (Eq.~\ref{fishermat}) as
discussed above.  The variance $\Delta\theta_{q}$ is thus a measure of
how sensitive the data set is to nonlinear scales; the larger the
variance, the less small--scale information the data contains.

Now, suppose that we compress all of the velocity information into a
single moment, $u= \sum_i b_{i}\ v_i$, where $b_{i}$ is a $1\times N$
set of coefficients.  The $1\times 1$ covariance matrix for $u$ will be
given by
\begin{equation}
{\tilde R} = \langle u^2\rangle =
\sum_{i,j} b_i\langle v_iv_j\rangle b_j = \sum_{i,j} b_iR_{ij}b_j\ .
\label{Rtilde}
\end{equation}
{}From the definition of the Fisher matrix (Eq.~\ref{fishermat}) and the
likelihood (Eq.~\ref{like}) , the $1\times 1$ Fisher matrix for the
compressed data takes the form
\begin{eqnarray}
{\tilde F_{qq}} &=& -\langle {\partial^2\over \partial\theta_q^2}\
\ln \left[\sqrt{(1/\tilde R)}\ \exp (-u^2/({2\tilde R})\right]\rangle
\nonumber\\ &=&
\langle {\partial^2\over \partial\theta_q^2}
\left({u^2\over 2\tilde R} + {\ln (\tilde R)\over 2}\right)\rangle
\nonumber\\ &=&{1\over 2} \langle u^2\rangle \left( {2\over \tilde R^3}
\left({\partial\tilde R\over\partial\theta_q}\right)^2 - {1\over\tilde
R^2}{\partial^2 \tilde R\over\partial\theta_q^2}\right) + \left(
{1\over\tilde R}{\partial^2\tilde R\over \partial\theta_q^2}- {1\over
\tilde R^2}\left({\partial\tilde R\over\partial\theta_q}
\right)^2\right)
\end{eqnarray}

It's convenient to normalize $b_i$ so that the moment $u$ has unit
variance, $\tilde R =\sum_{ij} b_iR_{ij}b_j= 1$; with this normalization
we can write
\begin{equation}
{\tilde F_{qq}}= {1\over 2}\left({\partial\tilde
R\over\partial\theta_q}\right)^2=\sum_{ij} {1\over 2}\left(b_i{\partial
R_{ij}\over\partial\theta_q} b_j\right)^2
\end{equation}

Note that this normalization will hold only for a particular covariance
matrix $R_{ij}$, and hence for a particular choice of parameters (see
Sec.~\ref{sec-anal})

Since the variance $\Delta\theta_{q}^{2}$ is inversely proportional to
$\tilde F_{qq}$, we can find the single moment that carries the minimum
information about $\theta_{q}$ by finding the $b_i$ which minimizes the
quantity on the right hand side, subject to the normalization
constraint.  This problem is solved by introducing a Lagrange multiplier
and extremizing the quantity
\begin{equation}
\sum_{ij} b_{i}{\partial R_{ij}\over
\partial \theta_{q}} b_{j}      - \lambda b_iR_{ij}b_j
\end{equation}
with respect to $b_{i}$, resulting in the equation
\begin{equation}
\sum_{j=1}^N \left({\partial R_{ij}\over \partial \theta_{q}}\right) b_j =
\sum_{j=1}^N\lambda R_{ij} b_j
\end{equation}

The equation above clearly represents an eigenvalue problem after
multiplying by $R^{-1}$ from the left; however, the matrix for this
eigenvalue problem is not normal.  Since the covariance matrix $R_{ij}$
is symmetric and positive definite, it can be Cholesky decomposed (see,
\eg, \citet{NR}) so that $R_{ij} =\sum_{p=1}^N L_{ip}L_{jp}$ for some
invertible matrix $L_{ij}$.  Plugging this into the equation above,
multiplying both sides by $L^{-1}_{ki}$ and summing over $i$ we obtain
an eigenvalue problem
\begin{equation}
\sum_{i,j,m}    \left( L^{-1}_{ki}{\partial R_{ij}\over
\partial \theta_{q}}L^{-1}_{lj}\right) \left(L_{ml}b_{m}\right) =
\sum_j \lambda \left( L_{jk}b_{j}\right)
\end{equation}
for the real, symmetric matrix ${\bf L}^{-1}\left({\partial {\bf R}/
\partial \theta_{q}}\right)\left({\bf L}^{-1}\right)^{T}$.   
Solving this eigenvalue problem gives us a set of $N$ orthogonal
eigenvectors $\sum_j L_{ji}(b_n)_j$ with corresponding eigenvalues
$\lambda_n$.  Each eigenvector has a corresponding moment $u_n= \sum_i
(b_n)_i v_i$.  The eigenvalue $\lambda_n$ of a moment $u_{n}$ is related
to the error bar $\Delta\theta_{q}$ that one could place on $\theta_{q}$
after compressing the velocities into that single moment, as can be seen
by manipulating the equations above:
\begin{equation}
{1\over
\Delta\theta_{q}}= \sqrt{F_{qq}} =
\sum_{i,j} {1\over\sqrt2}\ b_i{\partial R_{ij}\over
\partial
\theta_{q}} b_j= \sum_{i,j}{1\over\sqrt2}\ { b_i \left(\lambda R_{ij}b_j\right)} =
{\left|\lambda\right|\over\sqrt2}
\end{equation}
so that $\Delta\theta_{q} = \sqrt{2}/\left| \lambda\right|$.  Thus the
moment with the largest $|\lambda|$ is the moment that carries the
maximum possible amount of information about the parameter $\theta_{q}$.

The moments $u_{n}$ are statistically uncorrelated and of unit variance:
\begin{equation}
        \langle u_{n}u_{m}\rangle =\sum_{i,j} \langle
        (b_{n})_{i}v_{i}(b_{m})_{j}v_{j}\rangle =\sum_{i,j}
        (b_{n})_{i}R_{ij}(b_{m})_{j}=\sum_{i,p,j}
        (b_{n})_{i}L_{ip}L_{jp}(b_{m})_{j}= \delta_{nm}.
\end{equation}
Since we are assuming that the $v_{i}$, and thus the $u_{n}$, are
Gaussian random variables, this implies that the $u_{n}$ are
statistically independent, so that there is no overlap of information
among the $u_{n}$.  This suggests that if we convert the velocities
$v_{1}\ldots v_{N}$ into the $N$ moments
$u_{n}=\sum_{i}(b_{n})_{i}v_{i}=
\sum_i B_{ni}v_{i}$, there will be no loss of information, and the
transformation matrix $B_{ni}= (b_{n})_{i}$ will necessarily be
invertible.  Further, the statistical independence of the $u_{i}$
insures that if we compress the data by leaving out selected moments,
i.e. by keeping only selected rows of $B_{nj}$, the information
contained by those moments will be completely removed from the data.
The question becomes, then, how to select which moments to leave out.

\section{MOMENT SELECTION}
\label{sec-select}

If we order the moments $u_{n}$ in order of increasing eigenvalue,
\begin{equation}
        |\lambda_{1}|\le |\lambda_{2}|\le \ldots\le |\lambda_{N}|
\end{equation}
then we can interpret each moment as carrying successively more
information about $\theta_{q}$, with $u_{N}$ carrying the maximum
possible amount of information.  Since our goal is to produce a data set
that is less sensitive to the value of $\theta_{q}$ than the original
data, we should keep moments only up to some $N^{\prime}$, thus
discarding the moments that carry the most information about
$\theta_{q}$.  However, we would also like to keep as many moments as
possible in order to retain the maximum information about large scales.

In order to choose a value of $N^{\prime}$, we need to examine what
error bar $\Delta\theta_{q}$ we can put on the parameter $\theta_{q}$
using the compressed data.  Since the moments are independent, we can
write the Fisher matrix for the $N^{\prime}$ moments that were not
discarded as
\begin{equation}
\tilde F_{qq}= \sum_{n=1}^{N^{\prime}} \sum_{ij} {1\over 2}\left((b_n)_i{\partial
R_{ij}\over\partial\theta_q} (b_n)_j\right)^2=
\sum_{n=1}^{N^{\prime}}{1\over 2}\lambda_{n}^{2}
\end{equation}
so that the error bar that can be put on $\theta_{q}$ using the
compressed data is given by
\begin{equation}
        \Delta\theta_{q} = {1\over\sqrt{\tilde F_{qq}}}= \left[ {1\over
        2}\sum_{n=1}^{N^{\prime}}\ \lambda_{n}^{2}\right]^{-1/2}
\end{equation}
This result suggests that the number of moments kept, $N^{\prime}$,
should be chosen by adding up the sum of the squares of the smallest
eigenvalues until the desired sensitivity is reached.  For the purposes
of this paper, we shall adopt the following criterion: First, we
estimate the true size of the parameter $\theta_{q}=\theta_{qo}$ from
actual peculiar velocity data.  Then, we keep the largest number
$N^{\prime}$ moments that is still consistent with the requirement that
$\Delta\theta_{q}\ge \theta_{qo}$.  With this requirement, as long as
our estimate of the true value of $\theta_{q}$ is correct, our final set
of moments $u_{1}\ldots u_{N^{\prime}}$ will not contain enough
information to distinguish the value of $\theta_{q}$ from zero.

\section{POWER SPECTRUM MODEL}
\label{sec-pow}

In order to carry out the program outlined above, we need to construct a
model of the power spectrum such that the power on nonlinear scales is
proportional to a single parameter $\theta_{q}$.  To this end, we adopt
a model for the power spectrum of the form
\begin{equation}
        P(k)= P_{l}(k) + \theta_{q}P_{nl}(k),
\end{equation}
where $P_{l}(k)=0$ for $k>k_{nl}$ and $P_{nl}(k)=0$ for $k<k_{nl}$.
Here $k_{nl}$ is taken to be the wavenumber of the largest scale for
which density perturbations have become nonlinear.  Traditionally, the
scale of nonlinearity is related to the radius $r_T$ of a top hat filter
that results in a density contrast of unity. This corresponds to a
wavenumber $k_{nl}=2/r_T=0.25h$Mpc$^{-1}$, using the value
$r_T=8h^{-1}$Mpc \citep{ms93}.  We prefer to use the smaller value of
$k_{nl}=0.2h$Mpc$^{-1}$ in order to ensure a better separation of linear
and nonlinear information.

For $P_{l}(k)$, we use the BBKS parameterization of the power spectrum
\citep{BBKS}
\begin{equation}
	P_{l}(k) = \sigma_8^2 C
	k\big\{1+\big[6.4(k/\Gamma)+3(k/\Gamma)^{1.5}
	+(1.7k/\Gamma)^2\big]^{1.13}\big\}^{-2/1.13}
\end{equation}
where $\Gamma$ parameterizes the ``shape'' of the power spectrum and the
overall normalization is determined by $\sigma_8$, the standard
deviation of density fluctuations on a scale of $8h^{-1}$Mpc.  The
constant $C$ is determined via the direct relation between $\sigma_8$
and the power spectrum.  For models where the total density parameter
$\Omega=1$, the shape parameter is related to the density of matter,
$\Gamma = \Omega_m h$; however, we typically choose to ignore this
relationship and treat $\Gamma$ as a free parameter which we vary
independently.

Since we are interested in reducing the sensitivity of the data to the
full range of nonlinear scales, we take $P_{nl}(k)$ to be constant in
the range of interest, $P_{nl}(k)= P_{o}$ for $k_{nl}<k<k_{c}$. While
the non--linear power spectrum would be more accurately approximated by
$P_{nl}(k)\propto k^{-1}$ \citep{km92}, this choice tends to emphasize
the importance of scales with wavenumber just larger than $k_{nl}$,
whereas we prefer to weight all nonlinear scales equally.  Our choice of
constant $P_{nl}(k)$ also has the benefit of simplifying our
calculations considerably.  We have introduced a largest wavenumber
$k_{c}$ for velocity modes to account for the fact that modes smaller
than the scale of perturbations out of which a galaxy or cluster forms
will not contribute to its velocity.  For the analysis of galaxy data we
have adopted a value of $k_c = 6.0h$Mpc$^{-1}$, although the results are
fairly insensitive to the exact value chosen. For simulated data, it is
important to consider the dynamic range of the simulation when choosing
the value of $k_c$.  The constant $P_{o}$ is set by the requirement that
the contribution of nonlinear scales to the line--of--sight velocity
dispersion, which is what we have called $\sigma_{*}$ above, should be
equal to the value estimated from the actual velocity data.  Thus our
``true'' value for $\theta_{q}$ will be $\theta_{qo}=1$.  We can express
$P_{o}$ in terms of $\sigma_{*}$ by noting the relationship between the
velocity power spectrum and the velocity dispersion:
\begin{equation}
	\sigma_{*}^2= 4\pi \int_{k_{nl}}^{k_{c}} P_{v}(k) k^{2}dk= 4\pi
	H^2 \Omega_o^{1.2}\ \int_{k_{nl}}^{k_{c}} P_{nl}(k) dk.
\end{equation}
Since we have taken $P_{nl}$ to be constant over the given interval in
$k$, the integral is trivial, giving the relationship
\begin{equation}
	P_{o}= {\sigma_{*}^2\over 4\pi H^2\Omega_o^{1.2}
	(k_{nl}-k_{c})}.
\end{equation}

The value of $\sigma_*$ used will depend on what type of data is being
analyzed as well as the value of $k_{nl}$.  For velocity data drawn from
a simulation, the value of $\sigma_*^2$ can be calculated directly by
subtracting the contribution to the theoretical dispersion found using
$P_l(k)$ from the measured velocity dispersion of objects in the
simulated catalog.  For real velocity data, one can obtain guidance for
the choice of $\sigma_*$ from an analysis of the RMS peculiar velocity
of the survey \citep{W97}.

\section{ANALYSIS}
\label{sec-anal}

In this section we describe how to apply the formalism outlined above in
order to analyze a peculiar velocity survey consisting of
line--of--sight peculiar velocities $v_{i}$ and positions $\bf r_{i}$
for a set of $N$ galaxies.  We first discuss the construction of the set
of moments $u_{1}\ldots u_{N^{\prime}}$ that are insensitive to
nonlinear scales.  We then show how these moments can be used to
evaluate the likelihood of particular cosmological models.

Our first step is to construct the covariance matrix $R_{ij}$ for the
velocities using Eqs.~(\ref{vo}--\ref{Rvij}).  However, in order to do
this we must specify values for the parameters in the power spectrum
$P(k)$.  For the purposes of finding the moments $u_{1}\ldots
u_{N^{\prime}}$, we choose $\Gamma=0.35$ and $\beta=\Omega^{0.6}\sigma_8=0.5$, the
values that we consider to be the ``best fit'' to a wide variety of
experimental data.  While this choice of parameters will effect the
specific choice of moments, it should not have a significant effect on
the likelihood results.  As for the choice of $\theta_{q}$, for
simplicity we will take $\theta_{q}=0$.  Since we seek moments that
cannot distinguish between the ``true'' value $\theta_{q}=1$ and
$\theta_{q}=0$, this should also not effect our results.  Once $R_{ij}$
is calculated, we Cholesky decompose it by finding the invertible matrix
$L_{ij}$ such that $R_{ij} = \sum_{p=1}^N L_{ip}L_{jp}$.

The derivative ${\partial R_{ij}\over\partial \theta_{q}}$ is also
required; for our model of the power spectrum it can be calculated in an
identical way to $R_{ij}$ except that we must replace $P(k)$ with
$P_{nl}(k)$.  Here the only parameter that must be set is the amplitude
$P_{o}$ of the nonlinear power spectrum discussed Sec.~\ref{sec-pow}
above.

The next step is to diagonalize the matrix $ \sum_{i,j} L^{-1}_{ki}
{\partial R_{ij}\over \partial \theta_{q}}L^{-1}_{lj}$.  This gives us a
set of eigenvalues $\lambda_{n}$ and eigenvectors
$\sum_{j}L_{ji}(b_n)_j$.  The moment coefficients $(b_{n})_{k}$ are
obtained from the eigenvectors by multiplying by $L^{-1}_{ik}$ and
summing over $i$.  We order the eigenvalues in order of increasing
magnitude $|\lambda_{1}|\le |\lambda_{2}|\le \ldots |\lambda_{N}|$, and
discard the moments with $n> N^{\prime}$, where $N^{\prime}$ is
determined by the condition that
\begin{equation}
\Delta\theta_{q}= \left({1\over 2}\sum_{n=1}^{N^{\prime}} 
\lambda_{n}^{2}\right)^{-1/2} \le 1
\label{criterion}
\end{equation}
as discussed in Sec.~\ref{sec-compress}.

The moments $u_{n}$ are derived for a single choice of power spectrum
parameters; however, we would like to use them to calculate the
likelihood of power spectrum models with a range of parameter values.
Since the moments are just linear combinations of the line--of--sight
velocities, we can in principle use them to evaluate the likelihood of
any given power spectrum model as long as we can calculate the
covariance matrix for the modes in the model.  It is important to note,
though, that the modes will in general not be statistically uncorrelated
or have unit variance for any choice of parameters except those from
which they were derived.

Given a model for the power spectrum $P(k)$, the covariance matrix
$\tilde R_{nm}$ for the moments $u_{n}$ can be calculated from the
covariance matrix for the velocities $v_{i}$ and positions $\bf r_{i}$,
\begin{equation}
\tilde R_{nm}= \langle u_{n}u_{m}\rangle =
\langle (b_{n})_{i}v_{i}(b_{m})_{j}v_{j}\rangle =
(b_{n})_{i}\langle v_{i}v_{j}\rangle (b_{m})_{j}= (b_{n})_{i}R_{ij}
(b_{m})_{j}
\label{Rnm}
\end{equation}
The likelihood for the given power spectrum is then given as usual by
\begin{equation}
L(u_{1},\ldots,u_{N^{\prime}}; P(k))= \sqrt{|\tilde R^{-1} |}exp(
-{u_{n}}\tilde R^{-1}_{nm}u_{m}/2)
\end{equation}
where the values of the moments are calculated from the velocities
$v_{i}$ through $u_{n}= (b_{n})_{i}v_{i}$ where we sum over repeated
indices.

By calculating the likelihood over a range of power spectrum parameters,
the maximum likelihood values of the parameters can be determined.  If
these values differ substantially from the initial ``guess'' used to
calculate the moments, then the moments should be recalculated using the
maximum likelihood values.  This process can be iterated until the
maximum likelihood values are close to the initial ``guess''; this
ensures that the moments are optimum near the peak of the likelihood
function.

Since our goal has been to reduce the sensitivity of our data to
nonlinear velocities, it is of interest to examine the contributions to
the individual moments $u_n$ from different scale modes. By substituting
Eq.~(\ref{Rnm}) in Eq.~(\ref{Rij}), the variance for a given moment
$u_{n}$ can be written as
\begin{equation}
\langle u_{n}u_{n}\rangle = \tilde R_{nn}=
(b_{n})_{i}R_{ij}(b_{n})_{j}=
(b_{n})_{i}R^{(v)}_{ij}(b_{n})_{j}+(b_{n})_{i}(b_{n})_{i}
(\sigma_{i}^{2}+ \sigma_{*}^{2}).
\end{equation}
The second term is the contribution to the moment due to noise.  The
first term is the contribution to the moment from the velocity field;
this term can be further expanded as
\begin{equation}
(b_{n})_{i}R^{(v)}_{ij}(b_{n})_{j}= {H^2f^2(\Omega_o)\over 2\pi^2}\int
P(k) (b_{n})_{i} W^2_{ij}(k)(b_{n})_{j}\ dk = {H^2f^2(\Omega_o)\over
2\pi^2}\int P(k) W^{2}_{n}(k)\ dk
\end{equation}
where we have defined the window function for the $n$th moment as
\begin{equation}
W^{2}_{n}(k)\equiv (b_{n})_{i} W^2_{ij}(k)(b_{n})_{j}
\label{Wsq} 
\end{equation}
where $W^{2}_{ij}(k)$ is given in Eq.~(\ref{Wtsq}) in the appendix.  The
window function $W^{2}_{n}(k)$ tell us the sensitivity of the moment
$u_{n}$ to the scale corresponding to the wave number $k$.  This gives
us a check on our method; ideally, the moments that we retain should
have window functions that are maximum at large scales and relatively
small in the region $k_{nl}\ge k \ge k_{c}$ (see Sec.~\ref{sec-pow}).
However, since we have chosen our moments by their insensitivity to
small scales, there is no guarantee that they will necessarily be
sensitive to large scales.  Indeed, we have found that in some cases a
small number of the modes found by this method turn out to be
insensitive to almost {\it all} scales.  This can occur when a mode is
either dominated by far away galaxies with large errors or by a close
pair of galaxies; a moment representing the difference of the velocities
of two closely spaced galaxies is sensitive only to scales which are
smaller than the separation.  Since the moments are normalized to have
unit variance, the ones with low signal to noise can be found by
examining the contribution to the variance of each moment from the noise
part of the covariance matrix; moments with a noise contribution above
some threshold can be discarded.

\section{RESULTS FROM SIMULATED CATALOGS}
\label{results}

One concern is that the same small--scale, nonlinear effects that we are
trying to remove can also lead to deviations from Gaussianity, which our
method does not account for.  While it is plausible that these
deviations are small enough in typical velocity surveys as to not
significantly bias the results, the only way to be sure about this is to
test the method on realistic simulated catalogs.

While a more complete testing of our method will be presented in a
subsequent paper, in this section we present some results from
applying our method to simulated catalogs that illustrate the effects
of small--scale, nonlinear power and how they are mitigated in our analysis.

For our testing we have chosen simulated catalogs with $\approx1000$
galaxies designed to mimic the characteristics of the SFI survey
\citep{dacosta95}.  The catalogs were drawn from a $256^3$ N--body PM
(particle mesh) simulation with $\Gamma = 0.25$ and $\beta =
\Omega^{0.6}\sigma_8 = 0.46$.  In these simulations, the box size was
taken to be $512$ Mpc and the Hubble constant $h=H/100$~km~s$^{-1}{\rm
Mpc}^{-1}=0.75$; thus the box size in redshift space corresponds to a
diameter of 38,400~km~s$^{-1}$.  Galaxies were identified in these
simulations and assigned physical properties.  To duplicate the
characteristics of the SFI survey, galaxies were ``observed'' by
applying the same selection criteria.  Realistic scatter was added to
galaxy properties that duplicates the 15-20\% relative error in the SFI
inferred distances.  Finally, following Freudling {\it et al.} (1995) we
applied an inhomogeneous Malmquist correction to our catalogs.

\begin{figure}
\plotone{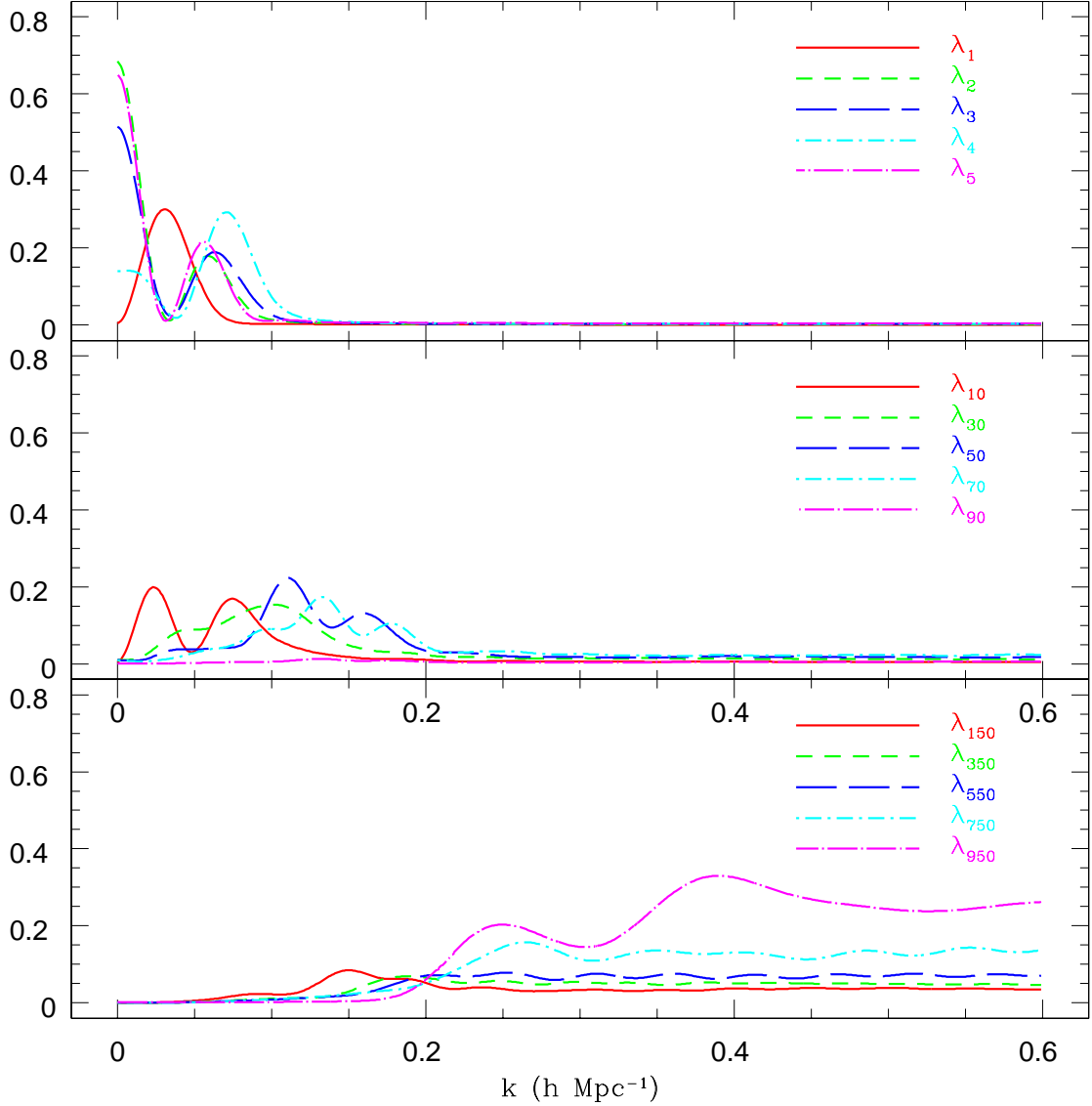}
\caption {The window functions in units of $6\cdot10^{-5}$. The
top panel shows the window functions associated with the five smallest
eigenvalues. The center panel shows the window functions associated with
five somewhat larger eigenvalues, while the bottom panel shows the
window functions selected from the entire range of eigenvalues. For each
panel we show the eigenvalue rank for each window function, as described
in Sec 4. It is clear that the window functions corresponding to lower
rank eigenvalues probe mostly large scales, whereas window functions
corresponding to large eigenvalues probe mostly nonlinear scales and
thus carry information that should not be used in an analysis based on
linear theory.
\label{winfun}}
\end{figure}

We performed the analysis described in Sec.~\ref{sec-anal} on these
simulated catalogs.  In Fig.~\ref{winfun} we show the window functions
for selected moments calculated for a typical catalog in order of
increasing eigenvalue, with the top plot showing the window functions
for the moments with the five lowest eigenvalues, the middle showing
five others associated with somewhat larger eigenvalues, and the bottom
plot showing five more selected from the whole range of eigenvalues.
This demonstrates that selecting moments that are least sensitive to
small scales does in fact generally result in moments that are most
sensitive to large scales; window functions of moments with larger
eigenvalues are successively larger on nonlinear scales as expected.
Thus the information contained in large eigenvalue moments comes mostly
from scales where fluctuations are nonlinear and should not be included
in a linear analysis.

\begin{figure}
\plotone{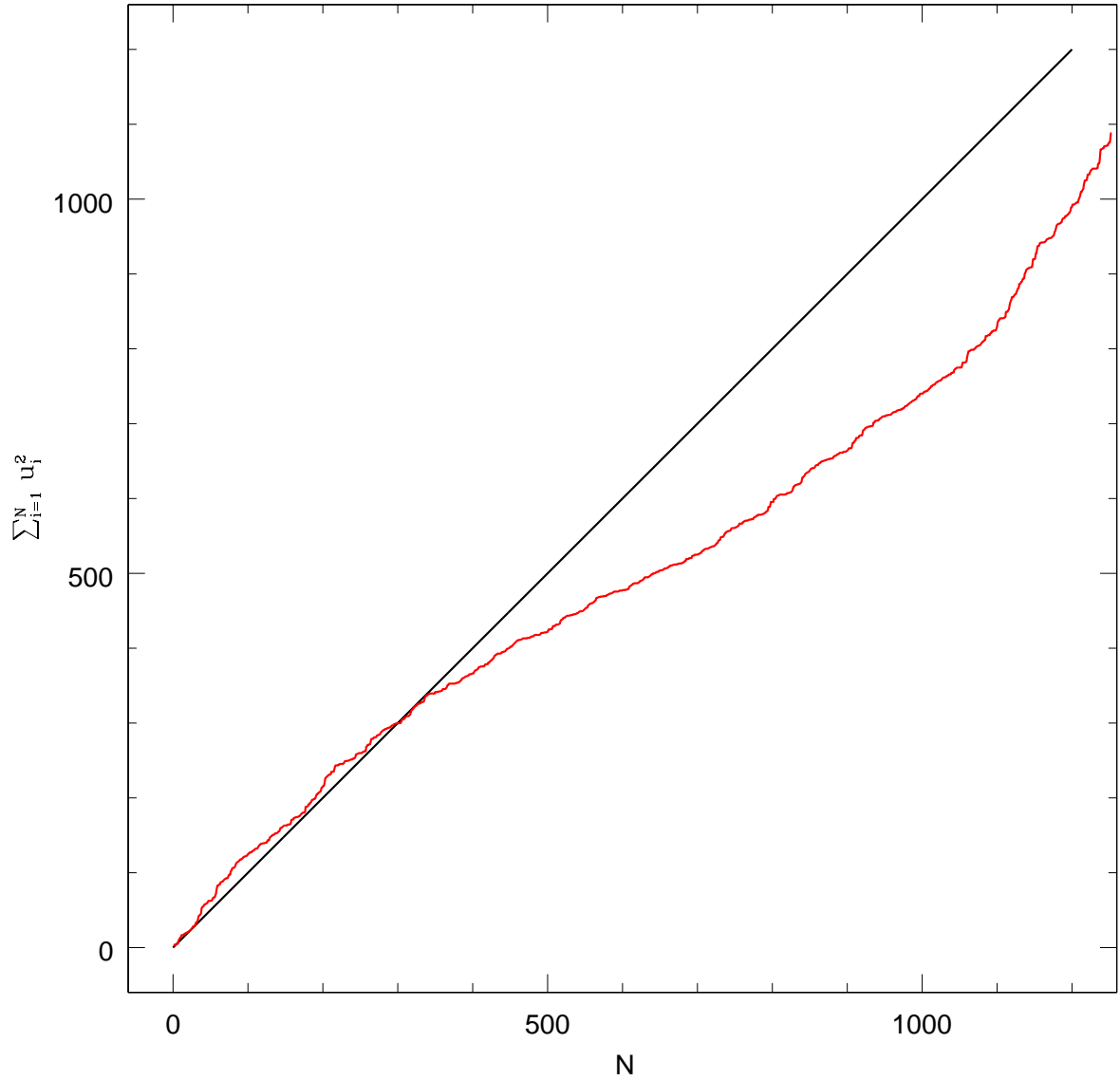}
\caption {The sum of the squares of the first $N$ moments versus
moment number $N$, where the moments are ranked in order of increasing
eigenvalue.  Note that for small $N$, the sum tracks a line with unit
slope, whereas for large $N$ the sum deviates from this line; this is an
indication that the non--linear effects are causing the large $N$ moments
to deviate from unit variance.
\label{ratio}}
\end{figure}

For our simulated catalogs, we know the ``true'' values of $\Gamma$ and
$\beta$.  If we use these true values as our ``guess'' (see
Sec.~\ref{sec-anal}) to calculate the optimum moments, then the values
of these moments calculated from the velocities should have unit
variance, since the power spectrum model should be an excellent fit to
the data.  However, non--linear effects can cause higher order moments to
deviate from unit variance.  In Fig.~\ref{ratio} we show the sum of the
first $N$ moments versus moment number $N$ for a typical catalog, where
the moments are ranked in order of increasing eigenvalue.  Note that for
small $N$, the sum tracks a line with unit slope, whereas for large $N$
the sum deviates from this line; this is an indication that the
non--linear effects are causing the large $N$ moments to deviate from
unit variance.

\begin{figure}
\plotone{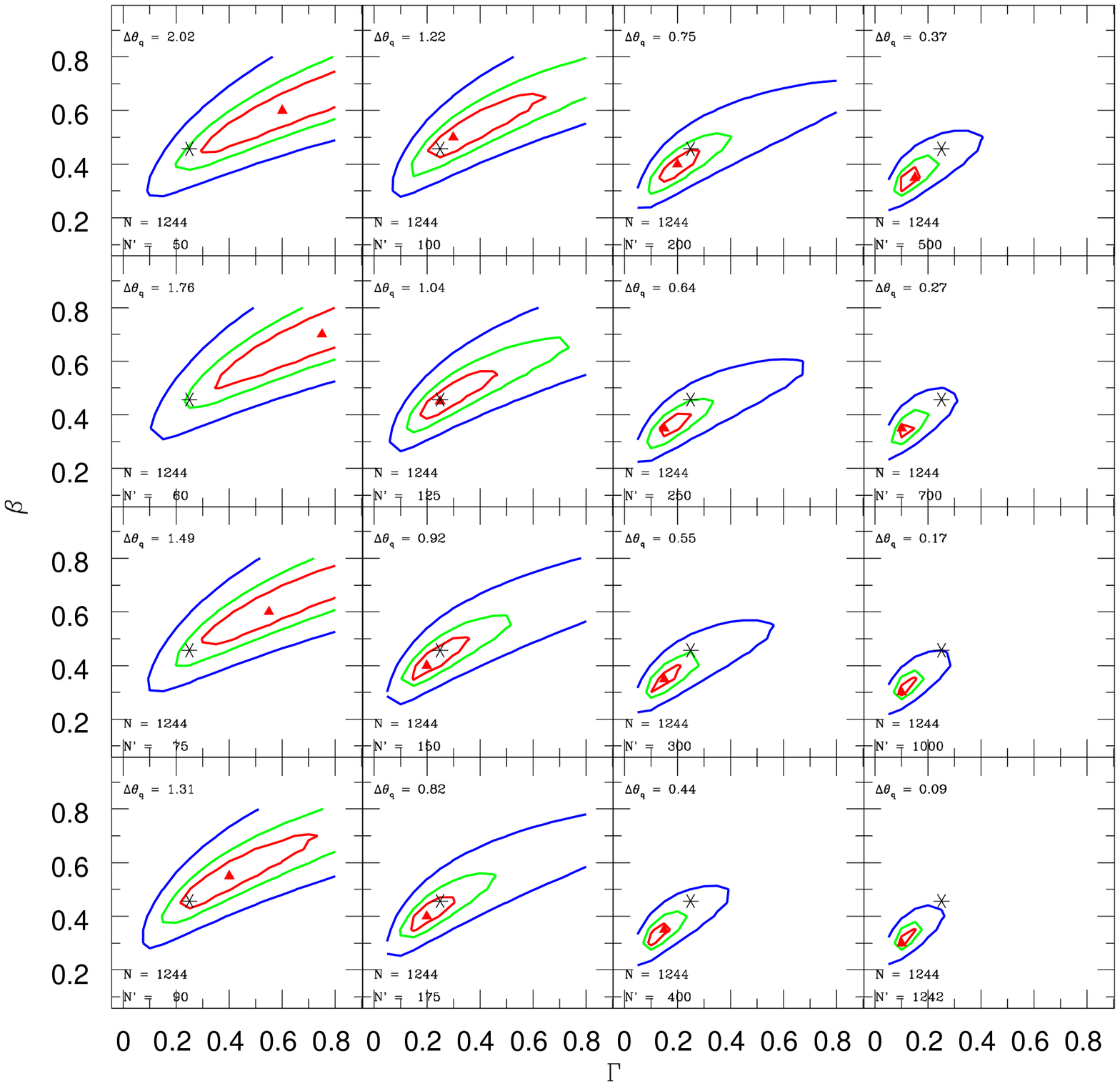}
\caption {Likelihood vs. $\Gamma$ and $\beta$ for the
simulated catalog as described in Sec.~\ref{results} for different
$N^{\prime}$, the number of moments kept; we also show the value of
$\Delta\theta_q$, the criterion for choosing $N^{\prime}$ as shown in
Eq.~(\ref{criterion}).  The panels show the maximum likelihood value
(solid triangles) and the contours corresponding to $0.5,0.1$ and $0.01$
of the maximum values. The asterisk in each panel is the ``true'' value
of $\Gamma=0.25$ and $\beta=0.457$ for the simulation. Increasing the
number of modes $N^\prime$ improves the accuracy of the maximum
likelihood values up to a point, but the inclusion of large eigenvalue
modes that carry information mostly from nonlinear scales can skew the
result away from the true value.
\label{maxval}}
\end{figure}

In Fig.~\ref{maxval} we show the results of the likelihood
analysis on a typical catalog for different number $N^{\prime}$
of moments kept.  For reference, we also give the value of
$\Delta\theta_q$ for each $N^{\prime}$ as discussed in
Sec.~\ref{sec-select}.  Here the closed triangles correspond to
the maximum likelihood values while the contours correspond to
$1/2$, $1/10$, and $1/100$ of the maximum likelihood.  The
asterisk symbol corresponds to the input values used for the
simulation, \ie the ``true'' values for $\Gamma$ and $\beta$.
We see that in this case, inclusion of all of the information
leads to the location of the maximum likelihood being skewed
away from the true values (see the panel with $N^{\prime}=N$).
However, when higher order moments are discarded, the location
of the maximum likelihood corresponds well with the true values.
For this particular catalog, with $\sigma_*=200$km/s, the
criterion of Eq.~(\ref{criterion}) would give $N^{\prime}\simeq
125$ for the optimum number of moments to keep.   The fact that
the discarding of higher order moments leads to a much better
agreement between the maximum likelihood location and the true
values is a good indication that our analysis method is
effectively removing small--scale, nonlinear velocity
information.    

Although the plots in Figs. 1--3 were calculated using a single catalog,
we note that the results from other catalogs that we analyzed were not
significantly different.

\section{DISCUSSION AND CONCLUSIONS}
\label{sec-conc}

In this paper we have presented a new method for the analysis of
peculiar velocity surveys which removes contributions to velocities from
small scale, nonlinear modes while retaining information about large
scale motions.  Our method selects a set of optimal moments constructed
as linear combinations of velocities which are minimally sensitive to
small scales.  We have shown how the overall sensitivity of a set of
moments to small scales can be quantified, and how to control this
sensitivity through the choice of the number of moments to retain.

As discussed above, the necessity of assuming Gaussian
statistics in our analysis raises the possibility that
deviations from Gaussianity caused by the collapse of
perturbations will interfere with the removal of small scale
power and introduce additional unpredictable biases.  While the
results of Sec. ~\ref{results} indicate that deviations from
Gaussianity are not having a large effect, careful testing of
our analysis method using simulated catalogs will be necessary to prove its
effectiveness at filtering small--scale power.  We are currently carrying
out tests using catalogs drawn from simulations with a variety of
parameter values, the results of which will be presented in a
subsequent paper.  This work will explore in more detail such issues as
moment selection, optimal values for constants to be used in the
analysis, and the dependence of the results of the analysis 
on whether the survey objects are galaxies or clusters of galaxies and
how these objects are selected.  We will also investigate 
differences in the small scale power present in simulations
produced using $PM$, $P3M$ and tree codes.   Once these tests
are completed, we plan to apply our formalism to analyze
existing velocity surveys, including the Mark III and SFI
catalogs.  

One of the merits of the formalism we have presented is its
versatility; it can be applied to a wide variety of surveys with
different geometries and densities.  The formalism also allows
for each object in the survey to have an independent velocity
error.  Versatility will be especially important as new distance
measurement techniques begin to produce large surveys that may
have a variety of characteristics.  Our formalism will be
particularly useful for surveys which use clusters of galaxies
as tracers of the velocity field, which are necessarily quite
sparse.

While we have focused on the use of our data
compression formalism for the determination of power spectrum
parameters through a likelihood analysis, it has broad
applicability as a data filtering technique.  Essentially, our
formalism ``rotates'' the vector consisting of survey object
velocities into a basis where the covariance matrix is diagonal
with the new moments ranked as to their sensitivity to small
scales.  Discarding moments containing small scale information
is equivalent to setting the value of these moments equal to
zero; the vector of moments can then be rotated back to the
survey object velocity basis.  The result is essentially a
``smoothed'' or ``linearized'' velocity data set, which can be used as input for
analysis methods that focus on large scale motions and assume
linear theory.  The amount and scale of the filtering can be
adjusted by varying the constants and thresholds used in the
construction of moments.

Finally, we note that the general technique we have developed for using
data compression to filter out unwanted information may be
useful in other areas of astrophysics; for example, in the
analysis of cosmic microwave background data.

\acknowledgments

We wish to thank Avishai Dekel and Ami Eldar for illuminating
conversations.  HAF and ALM wish to acknowledge support from the
National Science Foundation under grant number AST--0070702, the
University of Kansas General Research Fund and the National Center for
Supercomputing Applications.

\appendix

\section{APPENDIX}

The line--of--sight velocity ${\bf\hat r}_i\cdot{\bf v(r_i)}$ can be
written in terms of the Fourier transform of the velocity field
\begin{equation}
{\bf\hat r}_i\cdot{\bf v(r_i)}= {1\over\left( 2\pi\right)^3} \int
{\bf\hat r}_i\cdot{\bf\hat k}\ v({\bf k})e^{i{\bf k\cdot r_i}}
\end{equation}
The covariance matrix then becomes
\begin{eqnarray}
R^{(v)}_{ij}&=&\langle {\bf\hat r}_i\cdot{\bf v(r_i)}\ {\bf\hat
r}_j\cdot{\bf v(r_j)}\rangle\\ &=& {1\over \left(2\pi\right)^6}\int
d^3k\int d^3k^{\prime}\ \left( {\bf\hat r}_i\cdot{\bf\hat k}\right)\
\left({\bf\hat r}_j\cdot{\bf\hat k^{\prime}}\right)\
\langle v({\bf k})v^*({\bf k^{\prime}})\rangle\  e^{i({k\cdot \bf r_i-
k^{\prime}\cdot r_j})}\\ &=& {1\over \left(2\pi\right)^3}\int d^3k \
\left({\bf\hat r}_i\cdot{\bf\hat k}\right)\ \left({\bf\hat
r}_j\cdot{\bf\hat k}\right)\ P_v(k)e^{i {\bf k}\cdot ({\bf r_i -
r_j})}\\ &=& \int dk\ k^2\ P_v(k) W^2_{ij}(k)\\
\end{eqnarray}
where we have used the fact that $\langle v({\bf k})v^*({\bf
k^{\prime}})\rangle = P(k)\delta({\bf k-k^\prime})$ and we have defined
the tensor window function $W^2_{ij}(k)$ as the integral over the
possible directions of the vector ${\bf k}$,
\begin{equation}
W^2_{ij}(k) = {1\over 4\pi}\int d\Omega_k \left({\bf\hat
r}_i\cdot{\bf\hat k}\right)\ \left({\bf\hat r}_j\cdot{\bf\hat k}\right)\
e^{i {\bf k}\cdot ({\bf r_i - r_j})}
\label{Wtsq}
\end{equation}

In linear theory, the velocity power spectrum is related to the density
power spectrum by $P_{v}(k)= (H^2/k^2)f^2(\Omega_o)P(k)$, with
$f(\Omega_o)\approx
\Omega^{0.6}$.   This allows us to write $R^{(v)}$ as an
integral over the density power spectrum,
\begin{equation}
R^{(v)}_{ij} = {H^2f^2(\Omega_o)\over 2\pi^2}\int P(k)W^2_{ij}(k)\ dk,
\end{equation}


\begin{thebibliography}{}

\bibitem[Bardeen \etal (1986)]{BBKS} Bardeen, J. M., Bond, J. R., Kaiser,
N. \& Szalay, A. S. 1986 \apj\ 304 15
\bibitem[Croft \& Efstathiou(1994)]{C&E}Croft, R. \& Efstathiou, G., 1994.
Proceedings of the 11th Potsdam Cosmology Workshop : {\it Large Scale
Structure in the Universe}, ed. M\"{u}cket., J. P. et al, World
Scientific, astro--ph/9412024.
\bibitem[da Costa {\it et al.} (1995)]{dacosta95} da Costa, L.N.,
	Freudling, W., Wegner, G., Giovanelli, R., Haynes, M.P., \&
	Salzer, J.J.  1996, ApJL, 468, L5
\bibitem[Eldar (2000)]{eldar00}Eldar, A. 2000, PhD Thesis 
\bibitem[Feldman \& Watkins(1994)]{fw94} Feldman, H. A. \& Watkins, R. 1994 \apj\ 430 L17--20
\bibitem[Feldman \& Watkins(1998)]{fw98} Feldman, H. A. \& Watkins,
R. 1998 \apj\ 494 L129--132
\bibitem[Fisher(1935)]{fisher} Fisher, R. A. 1935, J. Roy. Stat. Soc., 98, 39 
\bibitem[Freudling \etal(1995)]{freudling95} Freudling, W., da Costa, L. N., Wegner, G.
  Giovanelli, R., Haynes, M. P., \& Salzer, J. J. 1995, AJ, 110, 2
\bibitem[Hoffman \& Zaroubi(2000)]{hz00} Hoffman, Y. \& Zaroubi, S.,
2000, \apj\ 535 L5
\bibitem[Hamilton(2000)]{H00} 2000 MNRAS 312 257--284
\bibitem[Hamilton \& Tegmark(2000)]{HT00} 2000 MNRAS 312 285--294
\bibitem[Jaffe \& Kaiser(1995)]{jk95}Jaffe, A. \& Kaiser, N. 1995 \apj\ 255 26
\bibitem[Kaiser(1988)]{Kaiser88} Kaiser, N. 1988 \mnras\ 231 149
\bibitem[Kendall \& Stuart(1969)]{KS} Kendall, M. G. \& Stuart, A. 1969 {\it The advanced Theory of
Statistics} Vol. 2, London: Grifin.
\bibitem[Kenney \& Keeping(1954)]{KK} Kenney, J.F., \& Keeping,
E.S. 1954, {\it Mathematics of statistics}, New York, Van Nostrand
company 3rd ed.
\bibitem[Klypin \& Melott(1992)]{km92} Klypin, A.A. \& Melott, A.L. 1992 \apj\ 399 397
\bibitem[Lauer \& Postman(1994)]{LP} Lauer, T. \& Postman, M. 1994 ApJ. 425 418--38
\bibitem[Melott \& Shandarin(1993)]{ms93} Melott, A. L. \& Shandarin,
S. F., 1993, \apj\ 410 469--481
\bibitem[Matsubara, Szalay \& Landy(2000)]{matsubara00} Matsubara, T., Szalay,
A.S. \& Landy, S.D., 2000, \apj\ 535:L1--L4
\bibitem[Press \etal(1992)]{NR} Press, W.H., Teukolsky, S.A.,
Vetterling, W.T. \& Flannery, B.P., 1992 {\it Numerical Recepies}
Cambridge University Press.
\bibitem[Riess, Press \& Kirshner(1995)]{RPK} Riess, A. G., Press,
W. H., \& Kirshner, R. P. 1995, ApJ, 438, L17
\bibitem[Silberman \etal(2001)]{silberman01} Silberman, L., Dekel, A.,
Eldar, A. \& Zehavi, I., 2001, Submitted to \apj\ , astro--ph/0101361
\bibitem[Sodre \& Lahav (1993)]{sodre93} Sodre Jr., L. \& Lahav, O. 1993, MNRAS, 260, 285
\bibitem[Tegmark, Taylor \& Heavens(1997)]{TTH} Tegmark, M.,
Taylor, A.N. \& Heavens, A.F., 1997, \apj\ 480 22--35
\bibitem[Watkins \& Feldman(1995)]{wf95} Watkins, R. \& Feldman, H. A. 1995 \apj\ 453 L73--76
\bibitem[Watkins(1997)]{W97} Watkins, R. 1997 \mnras\ 292 L59
\end{thebibliography}
\end{document}